\documentclass[sigconf, 10pt]{acmart}

\usepackage{booktabs} 

\setcopyright{rightsretained}
\usepackage{listings,xcolor}
\usepackage{subcaption}
\usepackage{enumitem}

\newcommand{\negspace}{\vspace{-0.5\baselineskip}}
\newcommand{\snegspace}{\vspace{-0.25\baselineskip}}
  
\setlist[itemize]{leftmargin=1em}
\newcommand{\customparagraph}[1]{{\bf\it #1}}

\copyrightyear{2019} 
\acmYear{2019}
\setcopyright{acmlicensed}
\acmConference[SIGMOD '19]{2019 International Conference on Management of Data}{June 30-July 5, 2019}{Amsterdam, Netherlands}
\acmBooktitle{2019 International Conference on Management of Data (SIGMOD '19), June 30-July 5, 2019, Amsterdam, Netherlands}
\acmPrice{15.00}
\acmDOI{10.1145/3299869.3314039}
\acmISBN{978-1-4503-5643-5/19/06}
    
\begin{document}

\title[FoundationDB Record Layer: A Multi-Tenant Structured Datastore]{FoundationDB Record Layer:\\ A Multi-Tenant Structured Datastore}
\newcommand{\apple}{Apple, Inc.}
\author{Christos~Chrysafis}
\affiliation{\apple}
\author{Ben~Collins}
\affiliation{\apple}
\author{Scott~Dugas}
\affiliation{\apple}
\author{Jay~Dunkelberger}
\affiliation{\apple}
\author{Moussa~Ehsan}
\affiliation{\apple}
\author{Scott~Gray}
\affiliation{\apple}
\author{Alec~Grieser}
\affiliation{\apple}
\author{Ori~Herrnstadt}
\affiliation{\apple}
\author{Kfir~Lev-Ari}
\affiliation{\apple}
\author{Tao~Lin}
\affiliation{\apple}
\author{Mike~McMahon}
\affiliation{\apple}
\author{Nicholas~Schiefer}
\affiliation{\apple}
\author{Alexander~Shraer}
\affiliation{\apple}

\begin{abstract}
The FoundationDB Record Layer is an open source library that provides a record-oriented data store with semantics similar to a relational database implemented on top of FoundationDB, an ordered, transactional key-value store. The Record Layer provides a lightweight, highly extensible way to store structured data. It offers schema management and a rich set of query and indexing facilities, some of which are not usually found in traditional relational databases, such as nested record types, indexes on commit versions, and indexes that span multiple record types. The Record Layer is stateless and built for massive multi-tenancy, encapsulating and isolating all of a tenant's state, including indexes, into a separate logical database. We demonstrate how the Record Layer is used by CloudKit, Apple's cloud backend service, to provide powerful abstractions to applications serving hundreds of millions of users. CloudKit uses the Record Layer to host billions of independent databases, many with a common schema. Features provided by the Record Layer enable CloudKit to provide richer APIs and stronger semantics with reduced maintenance overhead and improved scalability.
\end{abstract}

\maketitle

\section{Introduction}

Many applications require a scalable and highly available backend that provides durable storage. Developing and operating such backends presents many challenges. First, the high volume of data collected by modern applications, coupled with the large number of users and high access rates, requires smart partitioning and placement solutions for storing data to achieve \emph{horizontal scalability}. Modern storage systems must scale elastically in response to increases in user demand for both storage capacity and computation. Second, as systems become larger, formerly ``rare'' events such as network, disk, and machine failures become everyday occurrences. A core challenge of distributed systems is maintaining \emph{system availability and durability} in the face of these problems. Third, many techniques for addressing scalability and availability challenges, such as eventual consistency, create immense challenges for application developers. Implementing \emph{transactions} in a distributed setting remains one of the most challenging problems in distributed systems.

Fourth, as stateful services grow, they must support the needs of many diverse users and applications. This \emph{multi-tenancy} brings many challenges, including isolation, resource sharing, and elasticity in the face of growing load. Many database systems intermingle data from different tenants at both the compute and storage levels. Retrofitting resource isolation to such systems is challenging. Stateful services are especially difficult to scale elastically because state cannot be partitioned arbitrarily. For example, data and indexes cannot be stored on entirely separate storage clusters without sacrificing transactional updates, performance, or both.

These problems must be addressed by any company offering stateful services.
Yet despite decades of academic and industrial research, they are notoriously difficult to solve correctly and require a high level of expertise and experience. Big companies use in-house solutions, developed and evolved over many years by large teams of experts. Smaller companies often have little choice but to pay larger cloud providers or sacrifice durability.

FoundationDB~\cite{FDB} democratizes industrial-grade highly-available and consistent storage, making it freely available to anyone as an open source solution~\cite{FDBGithub} and is currently used in production at companies such as Apple, Snowflake, and Wavefront. While the semantics, performance, and reliability of FoundationDB make it extremely useful, FoundationDB's data model, an ordered mapping from binary keys to binary values, is often insufficient for applications. Many of them need structured data storage, indexing capabilities, a query language, and more. Without these, application developers are forced to reimplement common functionality, slowing development and introducing bugs.

To address these challenges, we present the FoundationDB Record Layer: an open source record-oriented data store built on top of FoundationDB with semantics similar to a relational database~\cite{recordLayerAnnouncement, recordLayerGithub}. The Record Layer provides schema management, a rich set of query and indexing facilities, and a variety of features that leverage FoundationDB's advanced capabilities. It inherits FoundationDB's strong ACID semantics, reliability, and performance in a distributed setting. These lightweight abstractions allow for multi-tenancy at an extremely large scale: the Record Layer allows creating isolated logical databases for each tenant---at Apple, it is used to manage billions of such databases---all while providing familiar features such as structured storage and transactional index maintenance.

The Record Layer represents structured values as Protocol Buffer~\cite{protobufs} messages called records that include typed fields and even nested records. Since an application's schema inevitably changes over time, the Record Layer includes tools for schema management and evolution. It also includes facilities for planning and efficiently executing declarative queries using a variety of index types. The Record Layer leverages advanced features of FoundationDB; for example, many aggregate indexes are maintained using FoundationDB's atomic mutations, allowing concurrent, conflict-free updates. Beyond its rich feature set, the Record Layer provides a large set of extension points, allowing its clients to extend its functionality even further. For example, client-defined index types can be seamlessly ``plugged in'' to the index maintainer and query planner. Similarly, record serialization supports client-defined encryption and compression algorithms. 

The Record Layer supports multi-tenancy at scale through two key architectural choices. First, the layer is completely stateless, so scaling the compute service is as easy as launching more stateless instances. A stateless design means that load-balancers and routers need only consider where the data are located rather than which compute servers can serve them. Furthermore, a stateless server has fewer resources that need to be apportioned among isolated clients. Second, the layer achieves resource sharing and elasticity with its \emph{record store} abstraction, which encapsulates the state of an entire logical database, including serialized records, indexes, and even operational state. Each record store is assigned a contiguous range of keys, ensuring that data belonging to different tenants is logically isolated. If needed, moving a tenant is as simple as copying the appropriate range of data to another cluster as everything needed to interpret and operate each record store is found in its key range. 

The Record Layer is used by multiple systems at Apple. We demonstrate the power of the Record Layer at scale by describing how CloudKit, Apple's cloud backend service, uses it to provide strongly-consistent data storage for a large and diverse set of applications~\cite{CloudKit}. Using the Record Layer's abstractions, CloudKit offers multi-tenancy at the extreme by maintaining independent record stores for each user of each application. As a result, we use the Record Layer on FoundationDB to host billions of independent databases sharing thousands of schemata. 
In the future, we envision that the Record Layer will be combined with other storage models, such as queues and graphs, leveraging FoundationDB as a general purpose storage engine to provide transactional consistency across all these data models. In summary, this work makes the following contributions:
 \begin{itemize}[noitemsep,topsep=3pt]
\item An open source layer on top of FoundationDB with semantics akin to those of a relational database.
\item The record store abstraction and a suite of techniques to manipulate it, enabling billions of logical tenants to operate independent databases in a FoundationDB cluster.
\item An extensible architecture allowing clients to customize core features including schema management and indexing.
\item A lightweight design that provides rich features on top of the underlying key-value store.
\end{itemize}

\negspace\snegspace
\section{Background on FoundationDB} \label{sec:fdb}

FoundationDB is a distributed, ordered key-value store that runs on clusters of commodity servers and provides ACID transactions over arbitrary sets of keys using optimistic concurrency control. Its architecture draws from the \emph{virtual synchrony} paradigm~\cite{gcs, Birman10virtuallysynchronous} whereby FoundationDB is composed of two logical clusters: one that stores data and processes transactions and another coordination cluster (running Active Disk Paxos~\cite{ActiveDiiskPaxos}) that maintains membership and configuration for the first cluster. This allows FoundationDB to achieve high availability while requiring only $F+1$ storage replicas to tolerate $F$ failures~\cite{Birman10virtuallysynchronous}. FoundationDB is distinguished by its deterministic simulation testing framework which can quickly simulate entire clusters under a variety of failure conditions in a single thread with complete determinism. In the past year, we have run more than 250 million simulations equivalent to more than 1870 years and 3.5 million CPU-hours. This rigorous testing in simulation makes FoundationDB extremely stable and allows its developers to introduce new features and releases in a rapid cadence, unusual among similar strongly-consistent distributed---or even centralized---databases.

\customparagraph{Layers.} Unlike most databases, which bundle together a storage engine, data model, and query language, forcing users to choose all three or none, FoundationDB takes a modular approach: it provides a highly scalable, transactional storage engine with a minimal yet carefully chosen set of features. For example, it provides no structured semantics, and each cluster provides a single, logical keyspace which it automatically partitions and replicates. \emph{Layers} can be constructed on top to provide various data models and other capabilities. Currently, the Record Layer is the most substantial layer built on FoundationDB.

\customparagraph{Transactions and Semantics.} FoundationDB provides ACID multi-key transactions with strictly-serializable isolation, implemented using multi-version concurrency control (MVCC) for reads and optimistic concurrency for writes. As a result, neither reads nor writes are blocked by other readers or writers. Instead, conflicting transactions fail at commit time and are usually retried by the client. Specifically, a client performing a transaction obtains a read version, chosen as the latest database commit version, by performing a \texttt{getReadVersion} (GRV) call and performs reads at that version, effectively observing an instantaneous snapshot of the database. Transactions that contain writes can be committed only if none of the values they read have been modified by another transaction since the transaction's read version.
Committed transactions are written to disk on multiple cluster nodes and then acknowledged to the client. FoundationDB executes operations within a transaction in parallel, while preserving the program order of accesses to each key and guaranteeing that a read following a write to the same key within a transaction returns the written value. FoundationDB imposes a 5 second transaction time limit, which the Record Layer compensates for using techniques described in Section~\ref{sec:overview}.

Besides create, read, update, and delete (CRUD) operations, FoundationDB provides atomic read-modify-write operations on single keys (e.g., addition, min/max, etc.). Atomic operations occur within a transaction like other operations, but they do not create read conflicts, so a concurrent change to that value would not cause the transaction to abort. For example, a counter incremented concurrently by many clients would be best implemented using atomic addition.

FoundationDB allows clients to customize the default concurrency control behavior by trading off isolation semantics to reduce transaction conflicts. \emph{Snapshot reads} do not cause an abort even if the read key was overwritten by a later transaction.
For example, a transaction may wish to read a value that is monotonically increasing over time just to determine if the value has reached some threshold.  In this case, reading the value at a snapshot isolation level (or clearing the read conflict range for the value's key) would allow the transaction to see the state of the value, but it would not result in conflicts with other transactions that may be modifying the value concurrently.

\customparagraph{Keys, values, and order.} Keys and values in FoundationDB are opaque binary values. FoundationDB imposes limits on key and value sizes (10kB for keys and 100kB for values) with much smaller recommended sizes (32B for keys and up to 10KB for values). Transactions are limited to 10MB, including the key and value size of all keys written and the sizes of all keys in the read and write conflict ranges of the commit. In production at Apple, with workloads generated by CloudKit through the Record Layer, the median and 99th percentile transaction sizes are approximately 7KB and 36KB, respectively. Keys are part of a single global keyspace, and it is up to applications to divide and manage that keyspace with the help of several convenience APIs such as the tuple and directory layers described next. FoundationDB supports range reads based on the binary ordering of keys. Finally, range clear operations are supported and can be used to clear all keys in a certain range or starting with a certain prefix.

\customparagraph{Tuple and directory layers.} Key ordering in FoundationDB makes tuples a convenient and simple way to model data. The tuple layer, included within the FoundationDB client library, encodes tuples into keys such that the binary ordering of those keys preserves the ordering of tuples and the natural ordering of typed tuple elements. In particular, a common prefix of the tuple is serialized as a common byte prefix and defines a key \emph{subspace}. For example, a client may store the tuple \texttt{(state,city)} and later read using a prefix like \texttt{(state,*)}.

The directory layer provides an API for defining a logical directory structure which maps potentially long-but-meaningful binary strings to short binary strings, reducing the amount of space used by keys. For example, if all of an application's keys are prefixed with its name, the prefix might be added to the directory layer. Internally, the directory layer assigns values to its entries using a sliding window allocation algorithm that concurrently allocates unique mappings while keeping the allocated integers small.

\negspace\negspace
\section{Design Goals and Principles} \label{sec:principles}

\begin{figure}[b]
\includegraphics[width=\columnwidth]{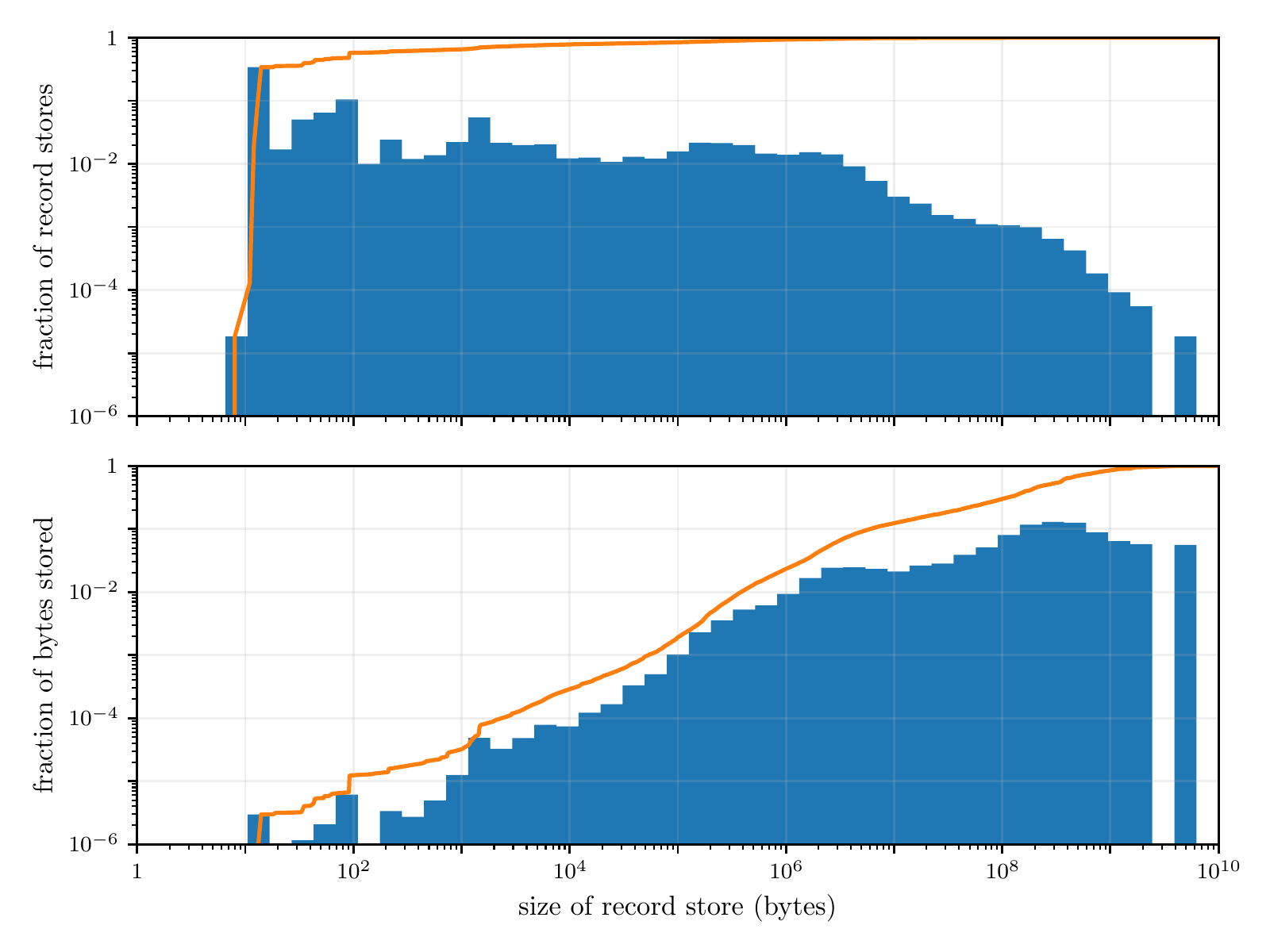}
\caption{\emph{Top:} The distribution of record store sizes for a 0.1\% sample of CloudKit-managed record stores as a normalized histogram (blue) and cumulative density function (orange). Notice that a substantial majority of record stores contain fewer than 1 kilobytes of record data.
\emph{Bottom:} The distribution of the total size of record stores of a particular size as a normalized histogram (blue) and cumulative density function (orange). The sample represents private databases and excludes many large record stores such as public databases. All sizes include primary record data and exclude metadata and index entries.}
\end{figure}\label{fig:recordstoresizes}

The Record Layer is designed to provide powerful abstractions for operations on structured data including transaction processing, schema management, and query execution. It aims to provide this rich set of features in a massively multi-tenant environment where it must serve the needs of a great number of tenants with diverse storage and retrieval needs. For example, some tenants store and locally synchronize a small amount of data while others store large amounts of data and query it in complex ways.

To illustrate this challenge concretely, Figure~\ref{fig:recordstoresizes} shows the distribution of the sizes of record stores used to store CloudKit's private databases \cite{CloudKit}. Although the vast majority of databases are quite small, much of the data is stored in relatively large databases. Furthermore, the data in the figure excludes larger databases such as the ``public'' databases shared across all users of an application, which can be terabytes in size.

In the face of such a range of database sizes, it is tempting to provision and operate separate systems to store ``small data'' and ``big data.'' However, such an approach poses both operational challenges and difficulties to application developers, who must contend with varying semantics across systems.
Instead, the Record Layer offers the semantics of a structured data store with massive multi-tenancy in mind by providing features that serve the diverse needs of its tenants and mechanisms to isolate them from each other.

\subsection{Design principles}

Achieving these goals required a variety of careful design decisions, some of which differ from traditional relational databases. This section highlights several of these principles.

\customparagraph{Statelessness.} In many distributed databases, individual database servers maintain ephemeral, server-specific state, such as memory and disk buffers, in addition to persistent information such as data and indexes. In contrast, the Record Layer stores all of its state in FoundationDB or returns it to the client so that the layer itself is completely stateless. For example, the Record Layer does not maintain any state about the position of a cursor in memory; instead, the context needed to advance a cursor in a future request is serialized and returned to the client as a continuation. 

The stateless nature of the layer has three primary benefits, which we illustrate with the same example of streaming data from a cursor. First, it simplifies request routing: a request can be routed to any of the stateless servers even if it requests more results from an existing cursor since there is no buffer on a particular server that needs to be accessed. Second, it substantially simplifies the operation of the system at scale: if the server encounters a problem, it can be safely restarted without needing to transfer cursor state to another server. Lastly, storing state in FoundationDB ensures that all state has the same ACID semantics as the underlying key-value store: we do not need separate facilities for verifying the integrity of our cursor-specific metadata. 

\customparagraph{Streaming model for queries.} The Record Layer controls its resource consumption by limiting its semantics to those that can be implemented on streams of records. For example, it supports ordered queries (as in SQL's \texttt{ORDER BY} clause) only when there is an available index supporting the requested sort order. This approach enables supporting concurrent workloads without requiring stateful memory pools in the server. This also reflects the layer's general design philosophy of preferring fast and predictable transaction processing over OLAP-style analytical queries.

\customparagraph{Flexible schema.} The atomic unit of data in the Record Layer is a \emph{record}: a Protocol Buffer~\cite{protobufs} message which is serialized and stored in the underlying key-value space. This provides fast and efficient transaction processing on individual records akin to a row-oriented relational database. Unlike record tuples in the traditional relational model, these messages can be highly structured; in addition to a variety of complex data types, these messages support nesting of record types within a field and repeated instances of the same field. As a result, list and map data structures can be implemented within a single record.
Because the Record Layer is designed to support millions of independent databases with a common schema, it stores metadata separately from the underlying data so that the common metadata can be updated atomically for all stores that use it.

\customparagraph{Efficiency.}
Implemented as a library rather than an independent client/server system, the Record Layer can be embedded in its client, imposing few requirements on how the actual server is implemented. Since FoundationDB is most performant at high levels of concurrency, nearly all of the Record Layer's operations are implemented asynchronously and pipelined where possible. We also make extensive use of FoundationDB-specific features, such as controllable isolation semantics, both within the layer's implementation and exposed to clients via its API. 

\customparagraph{Extensibility.} In the spirit of the FoundationDB's layered architecture, the Record Layer exposes a number of extension points through its API so that clients can extend its functionality. For example, clients can easily define new index types, methods for maintaining those indexes, and rules that extend its query planner to use those indexes in planning. This extensibility makes it easy to add features that are left out of the Record Layer's core, such as memory pool management and arbitrary sorting. Our implementation of CloudKit on top of the Record Layer (discussed in Section~\ref{sec:cloudkit}) makes substantial use of this extensibility with custom index types, planner behavior, and schema management.

\section{Record Layer Overview} \label{sec:overview}

\begin{figure}[t]
\includegraphics[scale=0.42]{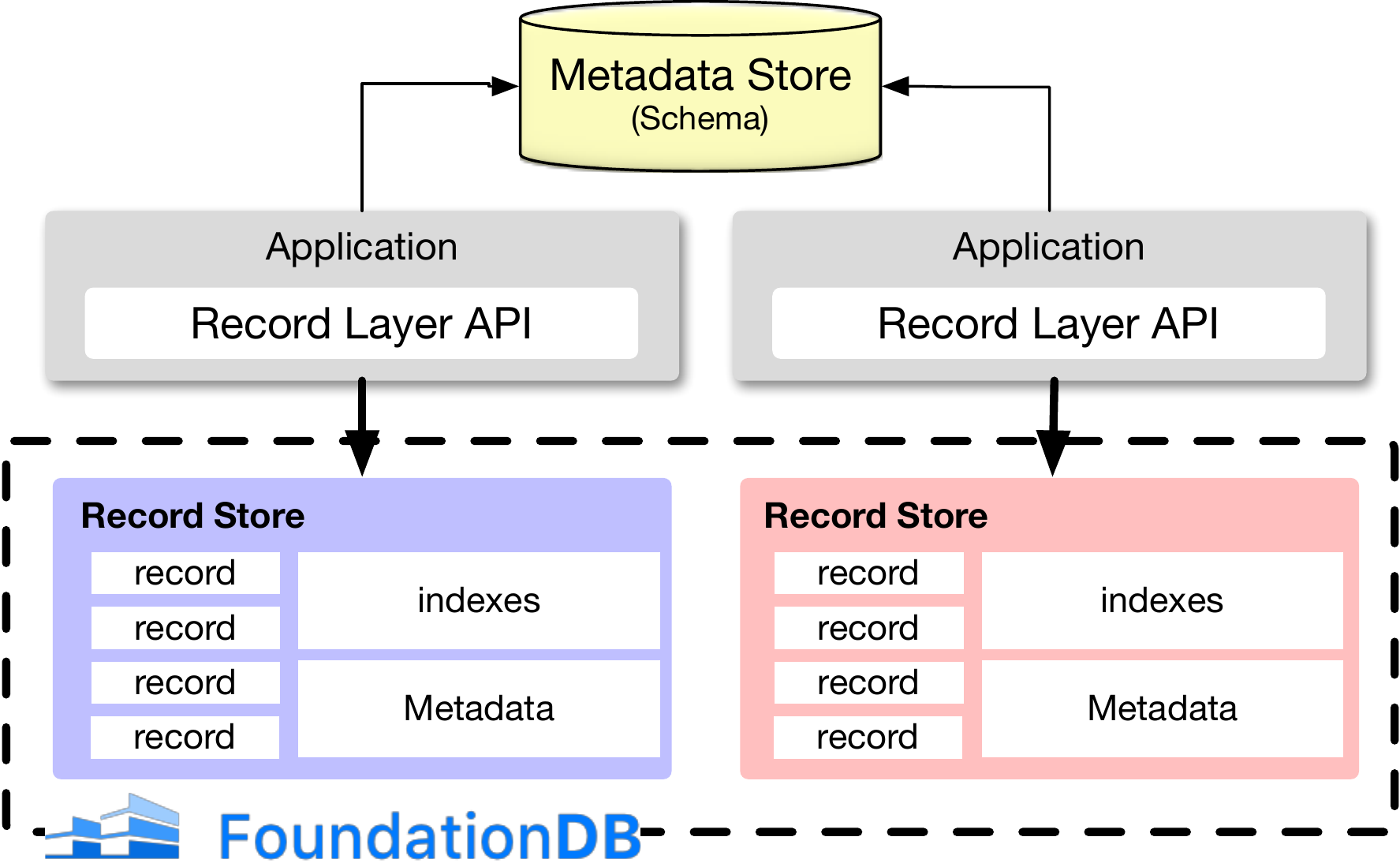}
\caption{Architecture of the Record Layer. The core \emph{record store} abstraction stores an entire logical database with a contiguous FoundationDB subspace, providing logical isolation between tenants.}
\negspace\negspace\snegspace
\label{fig:architecture}
\end{figure}

The Record Layer is primarily used as a library by stateless backend servers that need to store structured data in FoundationDB. It is used to store billions of logical databases, called \emph{record stores}, with thousands of schemata. Records in a record store are Protocol Buffer messages and a record type is defined with a Protocol Buffer definition. A record type resembles a table in a traditional relational database in that it defines the structure of multiple records of that type. However, unlike tables in a relational database, all record types within a record store are interleaved within the same extent. The schema, also called \emph{the metadata}, of a record store is a set of record types and index definitions on these types (see Section~\ref{sec:IndexDefs}). Metadata is versioned and may be stored in FoundationDB or elsewhere (metadata management and evolution is discussed in Section~\ref{sec:metadata}). The record store is responsible for storing raw records, indexes defined on the record fields, and the highest version of the metadata it was accessed with. This general architecture is shown in Figure~\ref{fig:architecture}.

Providing isolation between record stores is key for multi-tenancy. To facilitate resource isolation, the Record Layer tracks and enforces limits on resource consumption for each transaction, provides continuations to resume work, and can be coupled with external throttling. On the data level, the keys of each record store start with a unique binary prefix which defines a FoundationDB subspace. All the record store's data is logically co-located within the subspace and the subspaces of different record stores do not overlap. 

Primary keys and indexes are defined within the Record Layer using \emph{key expressions}, covered in detail in Appendix~\ref{sec:keyexp}.  A key expression defines a logical path through a record; applying it to a record extracts record field values and produces a tuple that becomes the primary key for the record or key of the index for which the expression is defined.  Key expressions may produce multiple tuples, allowing indexes to ``fan out'' and generate index entries for individual elements of nested and repeated fields. Since all record types are interleaved, both queries and index definitions may span all types of records in a record store.

To avoid exposing FoundationDB's limits on key and value sizes to clients, the Record Layer splits large records across a set of contiguous keys and splices them back together when deserializing split records. A special type of split, immediately preceding each record, holds the commit version of the record's last modification; it is returned with the record on every read.
The Record Layer supports pluggable serialization libraries, including optional compression and encryption of stored records.

The Record Layer provides APIs for storing, reading and deleting records, creating and deleting record stores and indexes in stores, scanning and querying records using the secondary indexes, updating record store metadata, managing a client application's directory structure, and iteratively rebuilding indexes when they cannot be rebuilt as part of a single transaction.

All Record Layer operations that provide a cursor over a stream of data, such as record scans, index scans, and queries, support continuations. A continuation is an opaque binary value that represents the starting position of the next available value in a cursor stream. Results are parceled to clients along with the continuation, allowing them resume the operation by supplying the returned continuation when invoking the operation again. This gives clients a way to control the iteration without requiring the server to maintain state, and it allows scan or query operations that exceed the transaction time limit to be split across multiple transactions.

The Record Layer leverages details of FoundationDB's implementation to improve data access speed. For example, record prefetching asynchronously preloads records into the FoundationDB client's read-your-write cache, but does not return them to the client application. When reading batches of many records, this can potentially save a context switch and record deserialization. The Record Layer also includes mechanisms to trade off consistency for performance, such as snapshot reads. Similarly, the layer exposes FoundationDB's ``causal-read-risky'' flag, which causes \texttt{getReadVersion} to be faster at the risk of returning a slightly stale read version during the rare case of a cluster reconfiguration. This is usually an acceptable risk; for example, ZooKeeper's ``sync'' operation behaves similarly~\cite{zookeeper}. Furthermore, transactions that modify state never return stale data since their reads are validated at commit time. Read version caching optimizes \texttt{getReadVersion} further by completely avoiding communication with FoundationDB if a read version was ``recently'' fetched from FoundationDB. Often, the client application provides an acceptable staleness and the last seen commit version as transaction parameters; the Record Layer uses a cached version as long as it is sufficiently recent and no smaller than the version previously observed by the client. This may result in reading stale data and may increase the rate of failed transactions in transactions that modify state. Version caching is most useful for read-only transactions that do not need to return the latest data and for low-concurrency workloads where the abort rate is low.

To help clients organize their record stores in FoundationDB's key space, the Record Layer provides a \texttt{KeySpace} API which exposes the key space in a fashion similar to a filesystem directory structure.  When writing data to FoundationDB, or when defining the location of a record store, a path through this logical directory tree may be traced and compiled into a tuple value that becomes a row key.
The \texttt{KeySpace} API ensures that all directories within the tree are logically isolated and non-overlapping. Where appropriate, it uses the directory layer (described in Section~\ref{sec:fdb}) to automatically convert directory names to small integers.

\negspace\negspace
\section{Metadata management} \label{sec:metadata}

The Record Layer provides facilities for managing changes to a record store's metadata. Since one of its goals is to support many databases that share a common schema, the Record Layer allows metadata to be stored in a separate keyspace from the data or even a separate storage system entirely. In most deployments, this metadata is aggressively cached by clients so that records can be interpreted without additional reads from the key-value store. 
This architecture allows low-overhead, per request, connections to a particular database.

\customparagraph{Schema evolution.} Since records are serialized into the underlying key-value store as Protocol Buffer messages (possibly after pre-processing steps, such as compression and encryption), some basic data evolution properties are inherited from Protocol Buffers: new fields can be added to a record type and appear as uninitialized in old records, and
new record types can be added without interfering with old records.
As a best practice, field numbers are never reused and should be deprecated rather than removed altogether.

The metadata is versioned in single-stream, non-branching, monotonically increasing fashion. Every record store tracks the highest version it has been accessed with in a small header within a single key-value pair. When a record store is opened, this header is read and the version compared with current metadata version.

Typically, the metadata will not have changed since the store was last opened, so these versions are the same.
When the version in the database is newer, a client has usually used an out-of-date cache to obtain the current metadata.
If the version in the database is older, changes need to be applied.
New records types and fields can be added by updating the Protocol Buffer definition.

\customparagraph{Adding indexes.} An index on a new record type can be enabled immediately since there are no existing records of that type.
Adding an index to an existing record type, which might already have records in the record store, is more expensive since it might require reindexing. Since records of different types may exist in the same key space, all the records need to be scanned when building a new index. If there are very few or no records, the index can be built right away within a single transaction. If there are many existing records, the index cannot be built immediately because that might exceed the 5 second transaction time limit. Instead, the index is disabled and the reindexing proceeds as a background job as described in Section~\ref{sec:IndexDefs}.

\customparagraph{Metadata versioning.} Occasionally, changes need to be made to the way that the Record Layer itself encodes data. For this, the same database header that records the application metadata version also records a storage format version, which is updated at the same time. Updating may entail reformatting small amounts of data or, in some cases, enabling a compatibility mode for old formats. We also maintain an ``application version'' for use by the client that can be used to track data evolution that is not captured by the metadata alone.
For example, a nested record type might be promoted to a top-level record type as part of data renormalization. The application version allows checking for these changes as part of opening the record store instead of implementing checks in the application code. If a series of such changes occur, the version can also be used as a counter tracking how far along we are in applying the changes to the record store.

\negspace
\section{Index Definition and Maintenance} \label{sec:IndexDefs}

Record Layer indexes are durable data structures that support efficient access to data, or possibly some function of the data, and can be maintained in a streaming fashion, i.e., updated incrementally when a record is inserted, updated, or deleted using only the contents of that record. Index maintenance occurs in the same transaction as the record change itself, ensuring that indexes are always consistent with the data. Our ability to do this efficiently relies heavily on FoundationDB's fast multi-key transactions. Efficient index scans use FoundationDB's range reads and rely on the lexicographic ordering of stored keys.
Each index is stored in a dedicated subspace within the record store so that indexes can be removed cheaply using FoundationDB's range clear operation.
Indexes may be configured with one or more \emph{index filters}, which allow records to be conditionally excluded from index maintenance, effectively creating a ``sparse'' index and potentially reducing storage space and maintenance costs.

\customparagraph{Index maintenance.} Defining an index type requires implementing an \emph{index maintainer} tasked with updating the index when records change. The Record Layer provides built-in index maintainers for a variety of index types (Section~\ref{sec:IndexTypes}). The index maintainer abstraction is directly exposed to clients, allowing them to define custom index types.

When a record is saved, we first check if a record already exists with the same primary key.
If so, registered index maintainers remove or update associated index entries for the old record, and the old record is deleted from the record store. A range clear to delete the old record is necessary as records can be split across multiple keys. Next, we insert the new record into the record store.
Finally, registered index maintainers insert or update any associated index entries for the new record. We use a variety of optimizations during index maintenance; for example, if an existing record and a new record are of the same type and some of the indexed fields are the same, the unchanged indexes are not updated.

\customparagraph{Online index building.} The Record Layer includes an online index builder used to build or rebuild indexes in the background. To ensure that an index is not used before it is fully built, indexes begin in a \emph{write-only} state where writes maintain the index but it cannot be used to satisfy queries. The index builder then scans the record store and instructs the index maintainer for that index to update the index for each encountered record. When the index build completes, the index is marked as \emph{readable}, the normal state where the index is maintained by writes and usable by queries. Index building is split into multiple transactions to reduce conflicts with concurrent mutations and avoid transaction size limits.

Indexes are defined by an index type and a \emph{key expression}, which defines a function from a record to one or more tuples consumed by the index maintainer and used to form the index key. The Record Layer includes a variety of key expressions (see Appendix~\ref{sec:keyexp}) and allows clients to define custom ones.

\negspace\negspace
\section{Index Types} \label{sec:IndexTypes}

The type of an index determines which predicates it can be used to evaluate. The Record Layer supports a variety of index types, many of which make use of specific FoundationDB features. Clients can define their own index types by implementing and registering custom key expressions and index maintainers (see Section~\ref{sec:IndexDefs}). In this Section, we outline  the \texttt{VALUE}, Atomic Mutation and \texttt{VERSION} indexes. Appendix~\ref{sec:RTIndexTypes} describes the \texttt{RANK} and \texttt{TEXT} index types, used for dynamic order statistics and full-text indexing respectively. 

Unlike in traditional relational systems, indexes can span multiple record types, in which case any fields referenced by the key expression must exist in all of the index's record types. Such indexes allow for efficient searches across different record types with a common set of search criteria.

\customparagraph{\texttt{VALUE} Indexes.} The default \texttt{VALUE} index type provides a standard mapping from index entry (a single field or a combination of field values) to record primary key. Scanning the index can be used to satisfy many common predicates, e.g., to find all primary keys where an indexed field's value is less than or equal to a given value.

\begin{figure*}[h]
\centering
\includegraphics[scale=0.35]{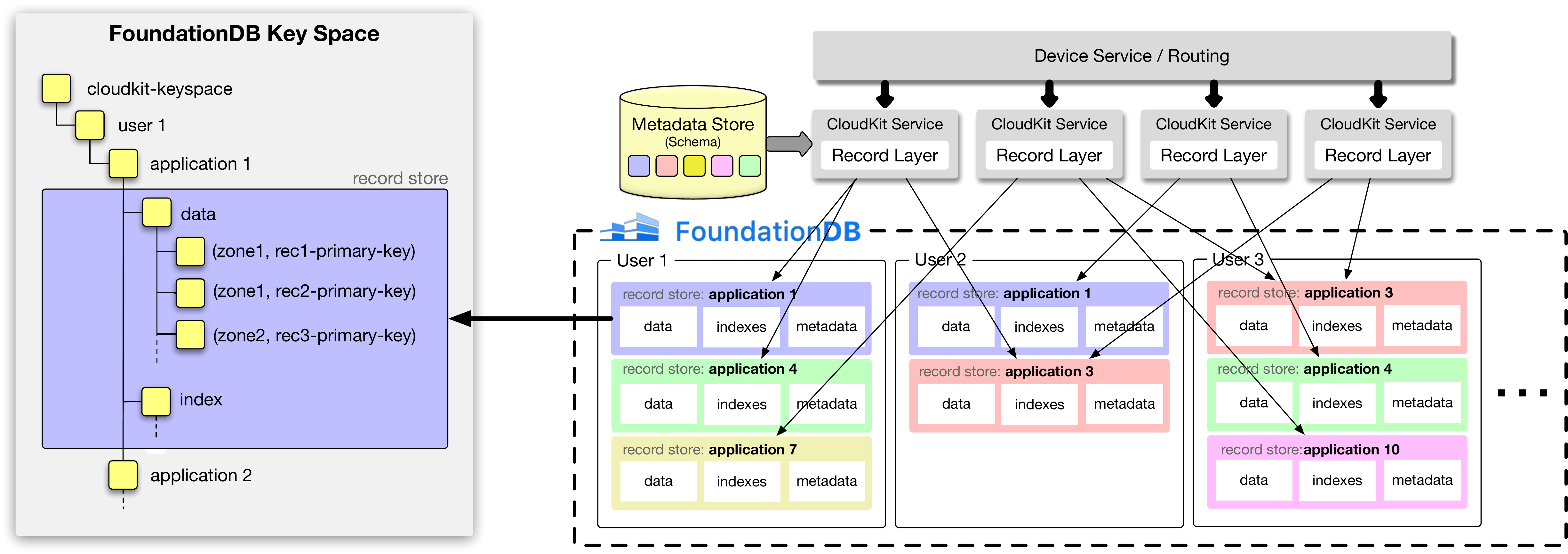}
\negspace\snegspace
\caption{CloudKit architecture using the Record Layer.}
\label{fig:cloudkit}
\negspace\snegspace
\end{figure*}

\customparagraph{Atomic mutation indexes.} Atomic mutation indexes are implemented using FoundationDB's atomic mutations, described in Section~\ref{sec:fdb}. These indexes are generally used to support aggregate statistics and queries. For example, the \texttt{SUM} index type stores the sum of a field's value over all records in a record store, where the field is defined in the index's key expression. In this case, the index contains a single entry, mapping the index subspace path to the sum value. The key expression could also include one or more grouping fields, in which case the index contains a sum for each value of the grouping field. While the maintenance of such an index could be implemented by reading the current index value, updating it with a new value, and writing it back to the index, such an implementation would not scale, as any two concurrent record updates would necessarily conflict. Instead, the index is updated using FoundationDB's atomic mutations (e.g., the \texttt{ADD} mutation for the \texttt{SUM} index), which do not conflict with other mutations.

The Record Layer currently supports the following atomic mutation index types, tracking different aggregate metrics:
\begin{itemize}[noitemsep,topsep=3pt]
\item \texttt{COUNT} - number of records
\item \texttt{COUNT UPDATES} - num.\ times a field has been updated
\item \texttt{COUNT NON NULL}- num.\ records where a field isn't null
\item \texttt{SUM} - summation of a field's value across all records
\item \texttt{MAX (MIN) EVER} - max (min) value ever assigned to a field, over all records, since the index has been created
\end{itemize}
Note that these indexes have a relatively low foot-print compared to \texttt{VALUE} indexes as they only write a single key for each grouping key, or, in its absence, a single key for each record store. However, a small number of index keys that need to be updated on each write can lead to high write traffic on those keys, causing high CPU and I/O usage for the FoundationDB storage servers that hold them. This can also result in increased read latency for clients attempting to read from these servers. 

\customparagraph{\texttt{VERSION}.} \texttt{VERSION} indexes are similar to \texttt{VALUE} indexes in that each has an index entry that maps to a primary key.
However, a \texttt{VERSION} index allows the index's key expression to include a special ``version'' field: a 12~byte value representing the commit version of the last update to the indexed record.
The version is guaranteed to be unique and monotonically increasing with time within a single FoundationDB cluster. The first 10~bytes are assigned by the FoundationDB servers upon commit, and only the last 2~bytes are assigned by the Record Layer, using a counter maintained by the Record Layer per transaction. Since versions are assigned in this way for each record insert and update, each record stored in the cluster has a unique version.

Since the version is only known upon commit, it is not included within the record's Protocol Buffer representation.
Instead, the Record Layer writes a mapping from the primary key of each record to its associated version in the keyspace adjacent to the records so that both can be retrieved efficiently with a single range-read.

Version indexes expose the total ordering of operations within a FoundationDB cluster. For example, a client can scan a prefix of a version index and be sure that it can continue scanning from the same point and observe all the newly written data. The following section describes how CloudKit uses this index type to implement change-tracking (sync).

\negspace\negspace
\section{Use Case: CloudKit} \label{sec:cloudkit}

CloudKit~\cite{CloudKit} is Apple's cloud backend service and application development framework, providing much of the backbone for storage, management, and synchronization of data across devices as well as sharing and collaboration between users. We describe how CloudKit uses FoundationDB and the Record Layer, allowing it to support applications requiring more advanced features such as the transactional indexing and query capabilities described in this paper.

Within CloudKit a given application is represented by a logical \emph{container}, defined by a schema that specifies the record types, typed fields, and  indexes that are needed to facilitate efficient record access and queries. The application clients store records within named zones.  \emph{Zones} organize records into logical groups which can be selectively synced across client devices.  

CloudKit assigns a unique FoundationDB subspace for each user, and defines a record store within that subspace for each application accessed by the user. This means that CloudKit is effectively maintaining $(\text{\# users}) \times (\text{\# applications})$ logical databases, each with their own records, indexes, and other metadata. CloudKit maintains billions of such databases. When requests are received from client devices, they are routed and load balanced across a pool of available CloudKit Service processes which access the appropriate Record Layer record store and service the request.

CloudKit translates the application schema into a Record Layer metadata definition and stores it in a metadata store (depicted in Figure~\ref{fig:cloudkit}). The metadata also includes attributes added by CloudKit such as system fields tracking record creation and modification time and the zone in which the record was written. The zone name is added as a prefix to primary keys, allowing efficient per-zone access to records. In addition to user-defined indexes, CloudKit maintains a number of ``system'' indexes, such as an index tracking the total record size by record type that is used for quota management.  

\negspace\negspace
\subsection{New CloudKit Capabilities}
CloudKit was initially implemented using Cassandra~\cite{cassandra} as the underlying storage engine. To support atomic multi-record operation batches within a zone, CloudKit uses Cassandra's light-weight transactions~\cite{cas}: all updates to the zone are serialized using Cassandra's compare-and-set (CAS) operations on a dedicated per-zone update-counter.  This implementation suffices for many applications using CloudKit, but it has two scalability limitations. First, there is no concurrency within a zone, even for operations making changes to different records. Second, multi-record atomic operations are scoped to a single Cassandra partition, which is limited in size; furthermore, Cassandra's performance deteriorates as the size of a partition grows. The former is a concern for collaborative applications, where data is shared among many users or client devices. These limitations require application designers to carefully model their data and workload such that records updated together reside in the same zone while making sure that zones do not grow too large and that the rate of concurrent updates is minimized. 

\begin{table}[t]
\begin{tabular}{l|c|c}
& Cassandra & Record Layer \\\hline
Transactions & Within Zone & Within Cluster \\\hline
Concurrency & Zone level & Record level \\\hline
Zone size limit & Cassandra & FoundationDB  \\
& partition size (GBs) & cluster size \\ \hline
Index Consistency & Eventual & Transactional \\ \hline
Indexes Stored & in Solr & in FoundationDB
\end{tabular}\caption{CloudKit on Cassandra and the Record Layer.}\label{tbl:cassandra}
\negspace\negspace\negspace\negspace\snegspace
\end{table}

The implementation of CloudKit on FoundationDB and the Record Layer addresses both issues, as summarized in Table~\ref{tbl:cassandra}. Transactions are scoped to the entire database, allowing CloudKit zones to grow significantly larger than before and supporting concurrent updates to different records within a zone. Leveraging these new transactional capabilities, CloudKit now exposes interactive transactions to its clients, specifically to other backend services that access CloudKit through gRPC~\cite{gRPC}. This simplifies the implementation of client applications and has enabled many new clients to use CloudKit.

Previously, only very few ``system'' indexes were maintained transactionally by CloudKit in Cassandra, whereas all user-defined secondary indexes were maintained in Solr. Due to high access latencies, these indexes are updated asynchronously, and queries that use them obtain an eventually consistent view of the data which requires application designers to work around perceived inconsistencies. With the Record Layer, user-defined secondary indexes are maintained transactionally with updates, so all queries return the latest data. 

\customparagraph{Personalized full-text search.} Users expect instant access to data they create such as emails, text messages, and notes. Often, indexed text and other data are interleaved, so transactional semantics are important. We implemented a personalized text indexing system using the \texttt{TEXT} index primitives described in Appendix~\ref{sec:RTIndexTypes} that now serves millions of users.
Unlike traditional search indexing systems, all updates are performed tranactionally, and no background jobs are needed to perform index updates and deletes.
In addition to providing a consistent view of the data, this approach also reduces operational costs by storing all data in one system.
Our system uses FoundationDB's key order to support prefix matching with no additional overhead and $n$-gram searches that require only $n$ key  entries
instead of the usual $O(n^2)$ keys needed to index all possible sub-strings.
The system also supports proximity and phrase search.

\customparagraph{High-concurrency zones.} With Cassandra, CloudKit maintains a secondary ``sync'' index from the values of the per-zone update-counter to changed records~\cite{CloudKit}. Scanning this index allows CloudKit perform a \emph{sync} operation that brings a mobile device up-to-date with the latest changes to a zone. The implementation of CloudKit using the Record Layer relies on FoundationDB's concurrency control and no longer maintains an update-counter that creates conflicts between otherwise non-conflicting transactions. To implement a sync index, CloudKit leverages the total order on FoundationDB's commit versions by using a \texttt{VERSION} index, mapping versions to record identifiers. To perform a sync, CloudKit simply scans the \texttt{VERSION} index.

However, versions assigned by different FoundationDB clusters are uncorrelated. This introduces a challenge when migrating data from one cluster to another, e.g., when moving users between clusters to improve load balancing and locality. The sync index must represent the order of updates across all clusters, so updates committed after the move must be sorted after updates committed before the move. CloudKit addresses this with an application-level per-user count of the number of moves, called the \emph{incarnation}. Initially, the incarnation is~1, and it is incremented each time the user's data is moved to a different cluster. On every record update, we write the user's current incarnation to the record's header; these values are not modified during a move.  The \texttt{VERSION} sync index maps (incarnation, version) pairs to changed records, sorting the changes first by incarnation, then version.

When deploying this implementation, we needed to handle existing data with an associated update-counter value but no version. Instead of using business logic to combine the old and new sync indexes, we used the \texttt{function} key expression (see Section~\ref{sec:keyexp}) to make this migration operationally straightforward, transparent to the application, and free of legacy code. Specifically, the \texttt{VERSION} index maps a \texttt{function} of the incarnation, version, and update counter value to a changed record, where the function is (incarnation, version) if the record was last updated with the new method and (0, update counter value) otherwise.
This maintains the order of records written using update counters and sorts all such records before records written with the new method.

\negspace\negspace
\subsection{Client Resource Isolation}

CloudKit services a very large number of concurrent requests, and it is vital that no one client or application overwhelms the system or impacts the performance of others. A number of Record Layer features were explicitly designed to ensure that individual requests consume a bounded amount of resources.

First, the Record Layer is designed to minimize the overhead of operations on top of the underlying key-value store. To illustrate this, we determined the median number of FoundationDB keys that were read or written while executing various common CloudKit operations. A query operation, which returns all records that match a given query, reads an average of $\sim$38.3 keys, of which $\sim$6.2 are not the records or index entries themselves, for a total overhead of $\sim$15\%. Simple requests for individual records are comparatively expensive, with an average of $\sim$13.3 key read of which $\sim$7.7 are not for record data. This reflects the general focus of CloudKit on providing higher level services, such as queries and synchronization over individual CRUD requests.

For record save operations, it is more complicated to properly estimate the overhead. For one, write overhead is dominated by the overhead of maintaining indexes, which depends on the number of indexes defined on the record's type. Furthermore, FoundationDB's commit time for write transactions does not have a simple relationship with the number of writes. The FoundationDB client buffers writes locally until commit time, when it ships all writes to the server along with the set of conflict ranges for the transaction. In practice, the write performance depends substantially on the number of conflicts produced rather than simply the number of writes. On average, a CloudKit transaction writes $\sim$8.5 records and makes $\sim$34.5 key writes associated with indexes, so the total index overhead is approximately $\sim$4 writes per record.

Today, the Record Layer does not provide the ability to perform in-memory query operations, such as hash joins, grouping, aggregation, or sorts. Operations such as sorting and joining must be assisted by appropriate index definitions. For example, efficient joins between records can be facilitated by defining an index across multiple record types on common field names. While this does impose some additional burden on the application developer, it ensures that the memory needed to complete a given request is limited to little more than the records accessed by the query. However, this approach may require a potentially unbounded amount of I/O to implement a given query. For this, we leverage the Record Layer's ability to enforce limits on the total number of records or bytes read while servicing a request.  When one of these limits is reached, the complete state of the operation is serialized and returned to the client as a continuation. The client may re-submit the operation with the continuation to resume the operation.
Because these operations are small, rate-based throttling mechanisms work more effectively.
With these limits, continuations, and throttling, we ensure that all clients make some progress even when the system comes under stress.

\section{Related Work} \label{sec:related}

Traditional relational databases offer many features including structured storage, schema management, ACID transactions, user-defined indexes, and SQL queries that make use of these indexes with the help of a query planner and execution engine.
These systems typically scale for read workloads but were not designed to efficiently handle transactional workloads on distributed data~\cite{Gray1996}.
 For example, in shared-nothing database architectures cross-shard transactions and indexes
 are prohibitively expensive and careful data partitioning, a difficult task for a complex application, is required. This led to 
 research on automatic data partitioning, e.g., ~\cite{Schism, autoPartitioning}. Shared-disk architectures are much more difficult to scale, 
 primarily due to expensive cache coherence and database page contention protocols~\cite{pdsFuture}.
 
 With the advent of Big Data, as well as to minimize costs, NoSQL datastores~\cite{dynamo, bigtable, pnuts, RiakKV, mongoDB, dynamoDB} offer the other end of the spectrum---excellent scalability but minimal semantics---typically providing a key-value API with no schema, indexing, transactions, or queries. As a result, applications needed to re-implement many of the features provided by a relational database. To fill the void, middle-ground ``NewSQL datastores'' appeared offering scalability as well as richer feature-sets and semantics~\cite{Spanner, cockroach, voltdb, cosmosDB, MemSQL, FDB}. For example, Google's Spanner~\cite{Spanner} supports ACID transactions, application-defined schema, and a SQL-based query language; secondary indexes and distributed queries were also added, more recently~\cite{SpannerSQL}. FoundationDB~\cite{FDB} takes a unique approach in the NewSQL space: it is highly scalable and provides ACID transactions, but it offers a simple key-value API with no built-in data model, indexing, or queries. This choice allowed FoundationDB to build a powerful, stable and performant storage engine without attempting to implement a one-size-fits-all solution. It was designed to be the foundation while layers built on top (implemented as client-side libraries) such as the Record Layer, provide higher-level abstractions. 

Multiple systems implement transactions on top of underlying NoSQL stores~\cite{wei2012cloudtps, Percolator, Tephra, omid, omidReloaded,megastore, warp, CockroachDB}. The Record Layer makes use of transactions exposed by FoundationDB to implement structured storage, complete with secondary indexes, queries, and more. Like Google Percolator~\cite{Percolator} the Record Layer is completely stateless, keeping all its metadata in the underlying FoundationDB data store.

The Record Layer has unique first-class support for multi-tenancy while most systems face the extreme challenge of retrofittig it.
Salesforce's architecture~\cite{salesforce} is similarly motivated by the need to support multi-tenancy within the database. For example, all data and metadata are sharded by application, and query optimization considers statistics collected per application and user. The Record Layer takes multi-tenancy support further through built-in resource tracking and isolation, a completely stateless design, and the record store abstraction. For example, CloudKit faces a dual multi-tenancy challenge as it services many applications, each with a very large user-base. Each record store encapsulates all of a user's data for one application, including indexes and metadata. This choice makes it easy to scale the system to billions of users by simply adding more clusters and moving record stores to balance the load and improve locality.

While many storage systems include support for full-text search, most provide this support using a separate system~\cite{CloudKit, RiakSolr, salesforce}, such as Solr~\cite{solr}, with eventual-consistency guarantees. In our experience with CloudKit, maintaining a separate system for search is challenging; it has to be separately provisioned, maintained, and made highly-available in concert with the database (e.g., with regards to fail-over decisions). MongoDB includes built-in support for full-text search, but indexes are not guaranteed to be consistent and so queries might not return all matching documents~\cite{mongoDBSearch}.

There is a broad literature on query optimization, starting with the seminal work of Selinger et al.~\cite{Selinger1979} on System R.
Since then, much of the focus has been on efficient search-space exploration. Most notably, Cascades~\cite{Graefe95thecascades} introduced a clean rule-based architecture for structuring query planners and proposed operators and transformation rules that are encapsulated as self-contained components. Cascades allows logically equivalent expressions to be grouped in the so-called Memo structure to eliminate redundant work. Recently, Greenplum's Orca query optimizer~\cite{Orca} was developed as a modern incarnation of Cascades' principles. We are currently in the process of developing an optimizer that uses the proven principles of Cascades, paving the way for the development of a full cost-based optimizer (Appendix~\ref{sec:query}).

\section{Lessons Learned} \label{sec:lessons}

The Record Layer's success at Apple validates the usefulness of FoundationDB's ``layer'' concept, where the core distributed storage system provides a semantically simple datastore upon which complex abstractions are built. This allows systems architects to choose the parts of the database that they need without working around abstractions that they do not. Building layers, however, remains a complex engineering challenge. To our knowledge, the Record Layer is deployed at a larger scale than any other FoundationDB layer. We summarize some lessons learned building and operating the Record Layer in the hope that they can be useful for both Record Layer adopters and developers of new layers.

\subsection{Building FoundationDB layers}

\customparagraph{Asynchronous processing to hide latency.} FoundationDB is optimized for throughput and not individual operation latencies, meaning that effective use requires keeping as much work outstanding as possible. Therefore, the Record Layer does much of its work asynchronously, pipelining it where possible. However, the FoundationDB client is single-threaded, with only a single network thread that talks to the cluster. Earlier versions of the Java bindings completed futures in the network thread, and the Record Layer used these for its asynchronous work, creating a bottleneck in that thread. By minimizing the amount of work done in the network thread, we were able to get substantially better performance and minimimze apparent latency on complex operations by interacting with the key-value store in parallel.

\customparagraph{Conflict ranges.} In FoundationDB, a transaction conflict occurs when some keys read by one transaction are concurrently modified by another. The FoundationDB API gives full control over these potentially overlapping read- and write-conflict sets.
One pattern is then to do a non-conflicting (snapshot) read of a range that potentially contains distinguished keys and adding individual conflicts for only these and not the unrelated keys found in the same range. Thus, the transaction depends only on what would invalidate its results. In the Record Layer, this technique is used for navigating the skip list used for the rank index described in Appendix~\ref{sec:RTIndexTypes}. Bugs due to incorrect manual conflict ranges are very hard to find, especially when mixed with business logic. For that reason, layers should generally define abstractions, such as indexes, for such patterns rather than relying on individual client applications to relax isolation requirements.

\negspace
\subsection{Using the Record Layer in practice}

\customparagraph{Metadata change safety.} The Protocol Buffer compiler generates methods for manipulating, parsing and writing messages, as well as static descriptor objects containing information about the message type, such as its fields and their types. These descriptors could potentially be used to build Record Layer metadata \emph{in code}. We do not recommend this approach over explicitly persisting the metadata in a metadata store, except for simple tests. One reason is that it is hard to atomically update the metadata code used by multiple Record Layer instances. For example, if one Record Layer instance runs a newer version of the code (with a newer descriptor), writes records to a record store, then an instance running the old version of the code attempts to read it, an authoritative metadata store (or communication between instances) is needed to interpret the data.
This method also makes it harder to check that the schema evolution constraints (Section~\ref{sec:metadata}) are preserved. We currently use descriptor objects to generate new metadata to be stored in the metadata store. 

\customparagraph{Relational similarities.} The Record Layer resembles a relational database but has sightly different semantics, which can surprise clients. For example, there is a single extent for all record types because CloudKit has untyped foreign-key references without a ``table'' association.
By default, selecting all records of a particular type requires a full scan that skips over records of other types or maintaining secondary indexes. 
For clients with a SQL-like table model, we now support emulating separate extents for each record type by adding a type-specific prefix to the primary key.

\snegspace
\subsection{Designing for multi-tenancy}

Multi-tenancy is remarkably difficult to add to an existing system. Hence, the Record Layer was built from the ground up to support massively multi-tenant use cases. We have gained substantial advantages from a natively multi-tenant design, including easier shard rebalancing between clusters and the ability to scale elastically. Our experience has led us to conclude that multi-tenancy is more pervasive than one would initially think. Put another way, many applications that do not explicitly host many different applications---as CloudKit does---can reap the benefits of a multi-tenant architecture by partitioning data according to logical ``tenants'' such as users, application functions, or some other entity.

\section{Future Directions}\label{sec:future}

The Record Layer's current feature set grew out of CloudKit's need to support billions of small databases, each with few users, and all with careful resource constraints. As databases grow in terms of data volume and number of concurrent users, the Record Layer may need to adapt and higher layers will be developed to support these new workloads and more complex query capabilities. However, we aim to ensure that the layer always retains its support for lightweight and efficient deployments. We highlight several future directions:

\customparagraph{Avoiding hotspots.} As the number of clients simultaneously accessing a given record store increases, checking the store header to confirm that the metadata has not changed may create a hot key. A general way to address hotspots is to replicate data at different points of the keyspace, making it likely that the copies are located on different storage nodes. For the particular case of metadata, where changes are relatively infrequent, we could also alleviate hotspots with caching. However, when the metadata changes, such caches need to be invalidated,  or, alternatively, an out-of-date cache needs to be detected or tolerated.

\customparagraph{Query operations.} Some query operations are possible with less-than-perfect indexing but within the layer's streaming model, such as a priority queue-based sort-with-small-limit or a limited-size hash join. For certain workloads, it may be necessary to fully support intensive in-memory operations with spill-over to persistent storage. Such functionality can be challenging at scale as it requires new forms of resource tracking and management and must be stateful for the duration of the query.

\customparagraph{Materialized views.} Normal indexes are a projection of record fields in a different order. \texttt{COUNT}, \texttt{SUM}, \texttt{MIN}, and \texttt{MAX} indexes maintain aggregates compactly and efficiently, avoiding conflicts by using atomic mutations. Adding materialized views, which can synthesize data from multiple records at once, is a natural evolution that would benefit join queries, among others. Adding support for materialized views to the key expressions API might also help the query planner reason about whether an index can be used to satisfy a query.

\customparagraph{Higher layers.} The Record Layer is close enough to a relational database that it could support a subset or variant of SQL, particularly once the query planner supports joins.
A higher level ``SQL layer'' could be implemented as a separate layer on top of the Record Layer without needing to work around choices made by lower-level layers. Similarly, a higher layer could support OLAP-style analytics workloads.

\snegspace
\section{Conclusions} \label{sec:conclusions}

The FoundationDB Record Layer offers rich features similar to those of a relational database, including structured schema, indexing, and declarative queries. Because it is built on FounationDB, it inherits its ACID transactions, reliability, and performance. The core \emph{record store} abstraction encapsulates a database and makes it easy to operate the Record Layer in a massively multi-tenant environment. The Record Layer offers deep extensibility, allowing clients to add features such as custom index types.
At Apple, we leverage these capabilities to implement CloudKit, which uses the Record Layer to offer new features (e.g., transactional full-text indexing), speed up key operations (e.g., with high-concurrency zones), and simplify application development (e.g., with interactive transactions) across billions of databases.

In building and operating the Record Layer at scale, we have made three key observations with broader applicability. First, the Record Layer's success validates FoundationDB's layered architecture in a large-scale system. Second, the Record Layer's extensible design provides common functionality while easily accommodating the customization needs of complex clients. It is easy to envision other layers that extend the Record Layer to provide rich, high-level functionality, as CloudKit does. Lastly, we find that organizing applications into logical ``tenants'', which might be users, features, or some other entity, is a powerful and practically useful way to structure a system and scale it to meet demand.

\snegspace
\begin{acks}
We thank Lyublena Antova and the anonymous reviewers for their insights and suggestions that helped us improve this paper. We are grateful to Mike Abbott, Bryan Davis, Eric Krugler, and Amol Pattekar for their guidance and support. Finally, we thank Ben Collins, Evan Tschannen and the FoundationDB team at Apple for their expertise.
\end{acks}

\clearpage
\bibliographystyle{abbrv}
\bibliography{record-layer}

\appendix

\section{Key Expressions} \label{sec:keyexp}

Indexes are defined by an index type and a key expression which defines a function from a record to one or more tuples consumed by the index maintainer and used to form the index key. The Record Layer includes a variety of key expressions and also allows clients to define custom ones.

\begin{figure}[b]
\negspace\negspace
\includegraphics[scale=0.6]{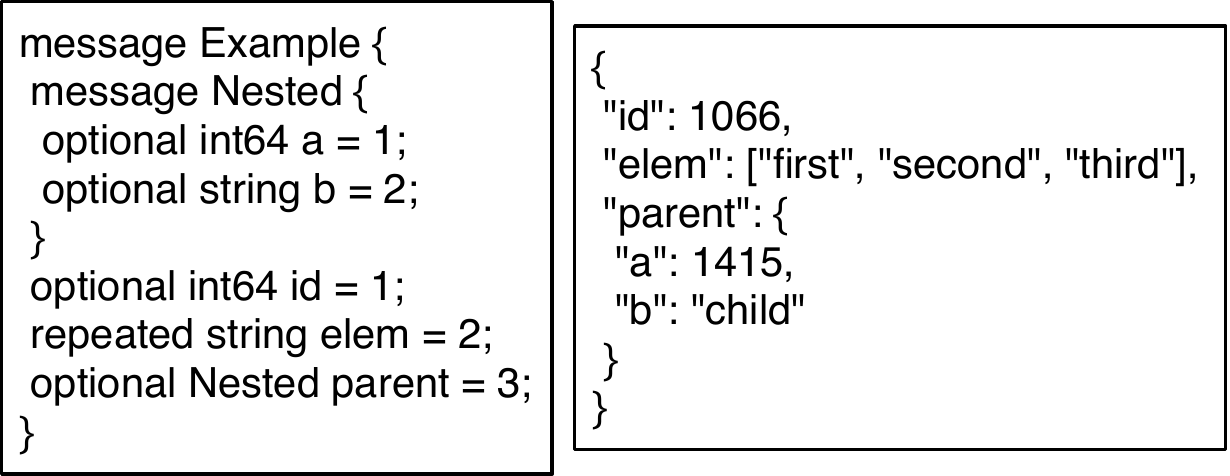}
\caption{Example record definition and an example.}
\label{proto}
\end{figure}

The simplest key expression is \texttt{field}. When evaluated against the  sample record in Figure~\ref{proto}, \texttt{field} \texttt{("id")} yields the tuple \texttt{(1066)}. Unlike standard SQL columns, Protocol Buffer fields permit repeated fields and nested messages. Nested messages can be accessed through the \texttt{nest} key expression. For example, \texttt{field("parent").nest("a")} yields \texttt{(1415)}. To support repeated fields, \texttt{field} expressions define an optional \texttt{FanType} parameter.
For a repeated field, \texttt{FanType} of \texttt{Concatenate} produces a tuple with one entry containing a list of all values within the field, while \texttt{Fanout} produces a separate tuple for each value. For example, \texttt{field("elem", Concatenate))} yields \texttt{(["first", "second", "third"])}, and \texttt{field("elem", Fanout))} yields three tuples: \texttt{("first"), ("second"), and ("third")}.

To create compound indexes, multiple key expressions can be concatenated. For example, \texttt{concat(field("id"), field("parent").nest("b"))} evaluates to the single tuple \texttt{(1066, "child")}. If a sub-expression produce multiple values, the compound expression will produce the tuples in the Cartesian product of the sub-expressions' values. 

The Record Layer also includes a variety of specialized key expressions. For example, the record type key expression produces a value that is guaranteed to be unique for each record type. When used in a primary key, this key expression allows users to treat record types like tables in a traditional relational database. It can also be used to satisfy queries about record types, such as a query for the number of records of each type that are stored within the database. Another special key expression, \texttt{version}, is described in Section~\ref{sec:IndexTypes}.

In addition to allowing client-defined key expressions, the Record Layer has \texttt{function} key expressions which allow for the execution of arbitrary, user-defined functions against records and their constituent fields. Function key expressions are very powerful and allow users to define custom sort orders for records, for example.
The layer also includes \texttt{groupBy} key expressions which can be used to divide an index into multiple sub-indexes and can be used for indexing aggregations like \texttt{SUM} and \texttt{COUNT}. The \texttt{KeyWithValue} key expression has two sub-express\-ions, one of which is included in the index entry's key and the other of which is in the entry's value. Such indexes are used as covering indexes that satisfy queries without resolving the indexed record.

\section{Rank and Text Index Types} \label{sec:RTIndexTypes}

\begin{figure}[t] 
\begin{subfigure}{\columnwidth} 
\includegraphics[width=3in]{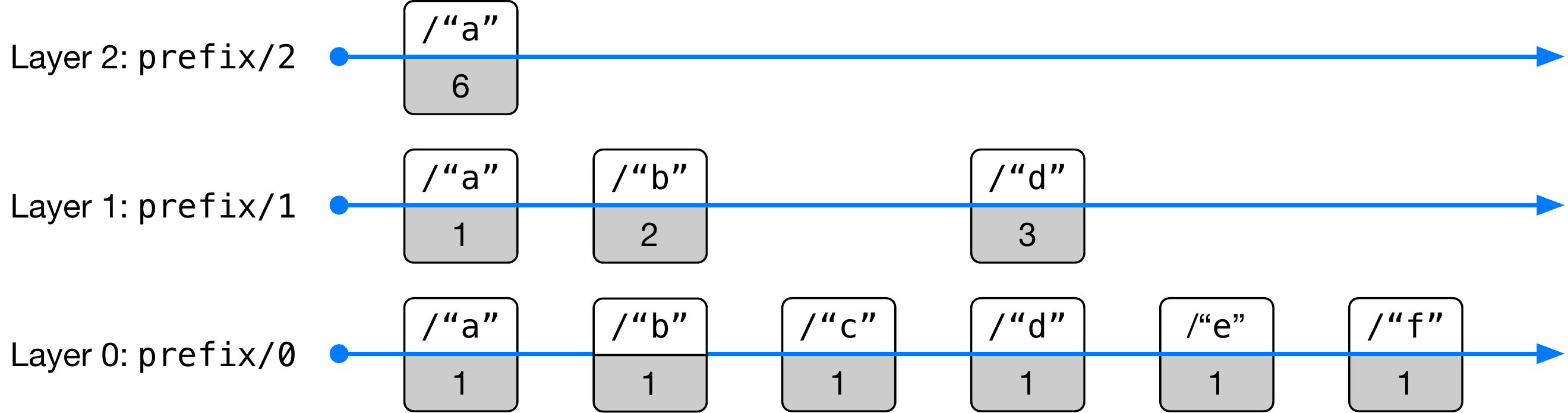}
\caption{Each contiguous key range used in by the skip-list is represented by a blue line and prefixed with the leading prefix of that subspace. Key-value pairs are shown as points on the keyspace line with a key and value (with a grey background).}
\end{subfigure}

\begin{subfigure}{\columnwidth} 
\includegraphics[width=3in]{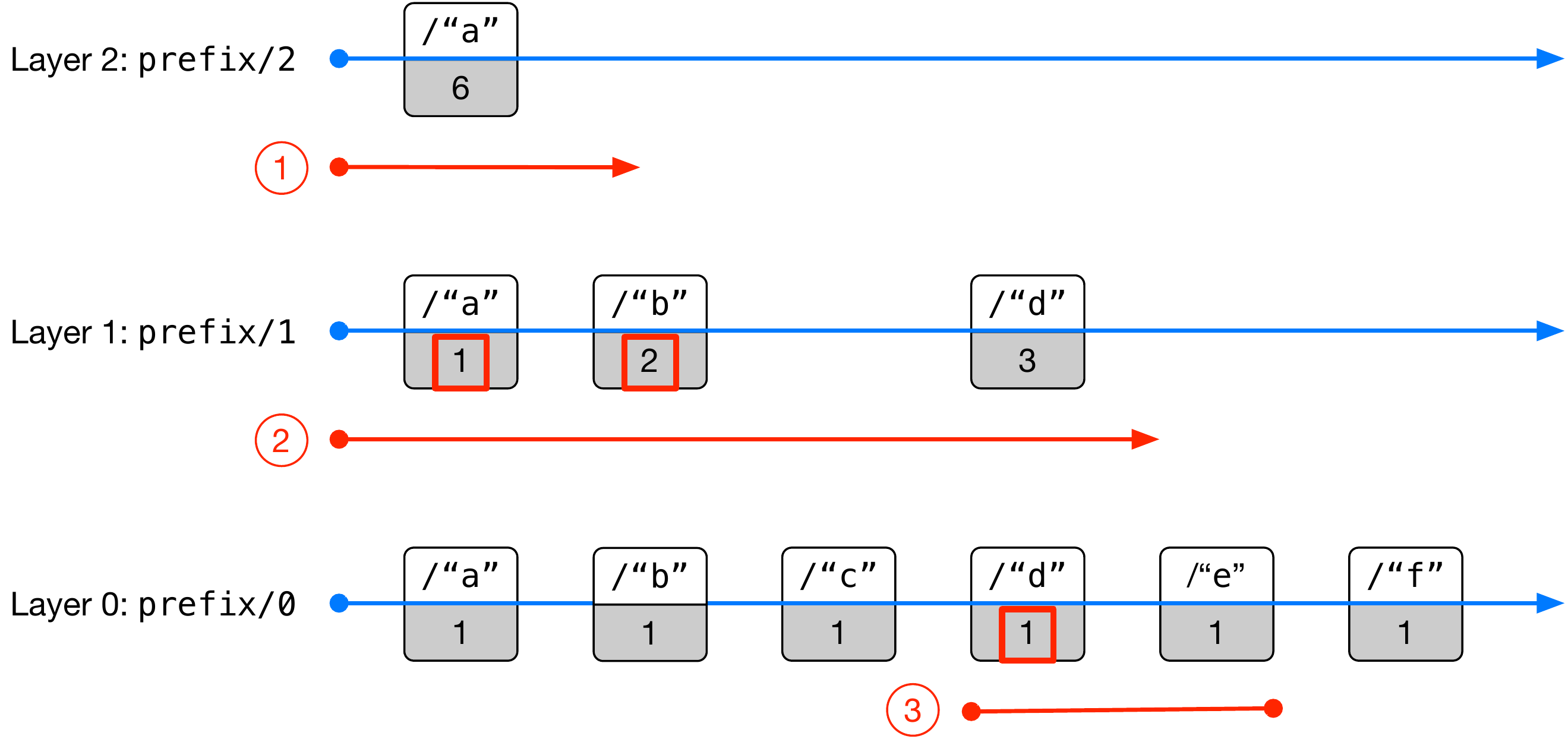}
\caption{An example of finding the rank for set element \texttt{"e"}. Scans of a range are shown with red arrows. (1) Scan the \texttt{prefix/2} subspace and find only \texttt{a}, which comes before \texttt{e}. (2) Scan scan the \texttt{prefix/1} subspace. Use the same-level fingers from \texttt{(prefix,1,"a")}, \texttt{(prefix,1,"b")}, contributing 1 and 2 to the sum, respectively. The last set member found is \texttt{d}, which also comes before \texttt{e}. (3) Scan the \texttt{prefix/0} subspace and find \texttt{e}. During our scan, use the same-level finger~\texttt{(prefix,0,"d")} which contributes 1 to the sum, yielding a total rank of 4 (the lowest ordinal rank is 0).}
\end{subfigure}
\caption{\snegspace
An example \texttt{RANK} index with 6 elements.}
\label{ranked-set}
\negspace\negspace\snegspace
\end{figure}

\customparagraph{\texttt{RANK} indexes.} The \texttt{RANK} index type provides efficient access to records by their ordinal rank (according to some key expression) and conversely to determine the rank of a field's value. For example, in an application implementing a leaderboard, finding a player's position in the leaderboard could be implemented by determining their score's rank using a \texttt{RANK} index. Another example is an implementation of a scrollbar, where data (e.g., query results) is sorted according to some field and the user can request to skip to the middle of a long page of results, e.g., to the $k$-th result. Instead of linearly scanning until the $k$-th result is found, using continuations to restart the scan if it runs too long, the client can query for the record with rank $k$ and begin scanning there.

Our implementation of the \texttt{RANK} index stores each index entry in a probabilistic augmented skip-list (Cormen et al.~\cite{clrs} describe a tree-based variant) persisted in FoundationDB such that each level has a distinct subspace prefix. Duplicate keys are avoided by attempting to read each key before inserting it. The lowest level of the skip-list includes every index entry, and each higher level contains a sample of entries in the level below it. For each level, each index entry contains the number of entries in the set that are greater or equal to it and less than the next entry in that level (all entries in the lowest level have the value 1). This is the number of entries skipped by following the skip-list ``finger'' between one entry and the next. In practice, an explicit finger is not needed: the sort order maintained by FoundationDB achieves the same purpose much more efficiently. Figure~\ref{ranked-set}(a) includes a sample index representing a skip-list with six elements and three levels.

To determine the ordinal rank of an entry, a standard skip-list search is performed, starting from the highest level. Whenever the search uses a finger connecting two nodes on the same level, we accumulate the value of the first node, i.e., the number of nodes being skipped. An example of this computation is shown in Figure~\ref{ranked-set}(b). The final sum represents the rank. Given a rank, a similar algorithm is used to determine the corresponding index entry. A cumulative sum is maintained and a range scan is performed at each level until following a finger would cause the sum to exceed the target rank, at which point the next level is scanned.

\customparagraph{\texttt{TEXT} indexes.} The \texttt{TEXT} index enables full-text queries on the contents of string fields. This includes simple token matching, token prefix matching, proximity search, and phrase search. The index stores tokens produced by a pluggable tokenizer using a text field as its input. Our inverted index implementation is logically equivalent to an ordered list of maps. Each map represents a postings list: it is associated with a token ($\text{token}_i$) and the keys in the map are the primary keys ($\text{pk}_j$) of records containing that token in the indexed text. Each value is a list of offsets in the text field containing that token expressed as the number of tokens from the beginning of the field. To determine which records contain a given token, a range scan can be performed on the index prefixed by that token to produce a list of primary keys, one for each record that contains that token. One can similarly find all records containing a token prefix. To filter by token proximity or by phrase, the scan can examine the relevant offset lists and filter out any record where the tokens are not within a given distance from each other or do not appear in the correct order.

To store the \texttt{TEXT} index, we use one key for each token/primary key pair, with the offset list in the value:\snegspace
\begin{lstlisting}[escapeinside=`', basicstyle=\footnotesize]
(prefix,token`$_1$',pk`$_1$')`$\rightarrow$'offsets`$_1$'
(prefix,token`$_1$',pk`$_2$')`$\rightarrow$'offsets`$_2$'
(prefix,token`$_2$' pk`$_3$')`$\rightarrow$'offsets`$_3$'
(prefix,token`$_3$',pk`$_4$')`$\rightarrow$'offsets`$_4$'
(prefix,token`$_3$',pk`$_5$')`$\rightarrow$'offsets`$_5$'
(prefix,token`$_3$',pk`$_6$')`$\rightarrow$'offsets`$_6$'
\end{lstlisting}
\snegspace
Note that the prefix is repeated in each key. While this is true for all indexes, the overhead is especially large for TEXT indexes due to the large number of entries. To address this, we reduce the number of index keys by ``bunching'' neighboring keys together so that for a given token, there might be multiple primary keys included in one index entry. Below is an example with a bunch size of 2, i.e., each index entry represents up to two primary keys.

\begin{lstlisting}[escapeinside=`', basicstyle=\footnotesize]
(prefix,token`$_1$',pk`$_1$')`$\rightarrow$'[offsets`$_1$',pk`$_2$',offsets`$_2$']
(prefix,token`$_2$',pk`$_3$')`$\rightarrow$'[offsets`$_3$']
(prefix,token`$_3$',pk`$_4$')`$\rightarrow$'[offsets`$_4$',pk`$_5$',offsets`$_5$']
(prefix,token`$_3$',pk`$_6$')`$\rightarrow$'[offsets`$_6$']
\end{lstlisting}
\snegspace

To insert a token \texttt{t} and primary key \texttt{pk}, we perform a range scan and find the biggest key $L$ that is less or equal to \texttt{(prefix,t,pk)} and the smallest key $R$ greater than \texttt{(prefix,} \texttt{t,pk)}. We then place a new entry in $L$' if the insertion does not cause this entry to exceed the maximum bunch size. If it does, the biggest primary key in the bunch is removed (this might be \texttt{pk} and its list of offsets) and is inserted in a new index key.
If the size of $R$'s bunch is smaller than the maximum bunch size, the bunch is merged with the newly created one. To delete the index entry for token \texttt{t} and primary key \texttt{pk}, the index maintainer performs a range scan in descending order from \texttt{(prefix,t,pk)}. The first key returned is guaranteed to contain the data for \texttt{t} and \texttt{pk}. If \texttt{pk} is the only key in the bunch, the entry is deleted. Otherwise, the entry is updated such that \texttt{pk} and its offsets are removed from the bunch, and if \texttt{pk} appears in the index key, the key is updated to contain the next primary key in the bunch.

Inserting an entry requires reading two key-value pairs and writing at most two (though usually only one). Deleting an entry requires reading and writing a single key-value pair. This access locality gives index updates predictable latencies and resource consumption. However, certain write patterns can result in many index entries where bunches are only partially filled. Currently, deletes do not attempt to merge smaller bunches together, although the client can request compactions.
Table~\ref{tab:moby_dick} shows a worked calculation of the possible space savings for this approach. To demonstrate space savings due to bunching, we used Melville's \emph{Moby Dick}~\cite{mobydick}, broken up into 233 $\sim$5 kilobyte documents. With whitespace tokenization, each document contains $\sim$431.8 unique tokens (so the primary key can be represented using 3~bytes) each appearing an average of $\sim$2.1 times within the document and with an average length of $\sim$7.8 characters. For the calculation, we use 10~bytes as the prefix size (smaller than the typical size we use in production). In practice, we find that this index required $\sim$4.9 kilobytes per document because not every bunch is actually filled. In fact, the average bunch size was $\sim$4.7, significantly lower than the maximum possible because some words appear rarely in the text--some even only once. To optimize further, we could bunch across tokens. An alternative would be to implement prefix compression in FoundationDB, but even then, there is per-key overhead in both the index and FoundationDB's internal B-tree data structure, so using fewer keys is still beneficial.

\begin{table}[]
    \begin{tabular}{r|c|c}
    & No bunch & Bunch size 20 \\ \hline\hline
    Prefix & 10 bytes & 10 bytes \\
    Token & 7.8 bytes & 7.8 bytes \\
    Primary key & 3 bytes & 3 bytes \\
    Encoding overhead & 2 bytes & 2 bytes \\ \hline
    Key size & 22.8 bytes & 22.8 bytes \\ \hline
    Offsets & 3 bytes & $\text{2 bytes} \times 20 = \text{40 bytes}$ \\
    Primary keys & --- & $\text{3 bytes} \times 19 = \text{57 bytes}$ \\ \hline
    Value size & 3 bytes & 97 bytes \\ \hline
    Total size / entry & 25.8 bytes & 119.8 bytes \\
    Approx. \# entries & 431.8 & $431.8 / 20 \approx 21.6$ \\ \hline
    Total size / document & 11.1 kB & 2.6 kB \\
    \end{tabular}
    \caption{Worked example illustrating the space savings provided by the bunched map.}
    \label{tab:moby_dick}
    \negspace\negspace\negspace
\end{table}

\section{Query planning and API} \label{sec:query}
The Record Layer has extensive facilities for executing declarative queries on a record store. While the planning and execution of declarative queries has been studied for decades, the Record Layer makes certain unusual design decisions.

\customparagraph{Extensible query API.}
The Record Layer has a fluent Java API for querying the database by specifying the types of records that should be retrieved, Boolean predicates that the retrieved records must match, and a sort order specified by a key expression (see Appendix~\ref{sec:keyexp}). Both filter and sort specifications can include ``special functions'' including aggregates, cardinal rank, and full-text search operations like $n$-gram and phrase search. This query language is akin to an abstract syntax tree for a SQL-like text-based query language exposed as an API, allowing consumers to directly interact with it in Java. Another layer on top of the Record Layer could provide translation from SQL or another query language.

\customparagraph{Query plans.} While declarative queries are convenient for clients, they need to be transformed into concrete operations such as index scans, union operations, and filters in order to execute the queries efficiently.
The Record Layer's query planner converts a declarative query specifying the records to return into an efficient combination of operations on the stream of records. The Record Layer exposes these query plans through the planner's API, allowing its clients to cache or otherwise manipulate query plans directly. This provides functionality similar to that of a SQL \texttt{PREPARE} statement but with the additional benefit of allowing the client to modify the plan if necessary~\cite{Orca}. Like SQL \texttt{PREPARE} statements, queries (and thus, query plans) may have bound static arguments (SARGS). In CloudKit we have leveraged this functionality to implement CloudKit-specific planning behavior by combining multiple plans produced by the Record Layer and binding the output of one plan to an argument of another.

\customparagraph{Cascades-style planner.}
We are currently evolving the Record Layer's planner from an ad-hoc architecture to a Cascades-style rule-based planner~\cite{Graefe95thecascades}. This new design supports deep extensibility, including custom planning logic defined completely outside the Record Layer by clients. Internally, we maintain a tree-structured intermediate representation (IR) of partially-planned queries that includes a variety of \emph{expressions}, including both logical operations (such as a sort order needed to perform an interaction of two indexes) and physical operations (such as scans of indexes, unions of streams, and filters). We implement the planner's functionality through a rich set of planner rules, which match to particular structures in the IR tree, optionally inspect their properties, and produce equivalent expressions.

Rules are automatically selected by the planner, but can be organized into ``phases'': for example, it is better to scan part of an index than to filter all records. They are also meant to be modular; several planner behaviors are implemented by multiple rules acting in concert. While this architecture make the code base easier to understand, its key benefit is allowing complex planning behavior by combining the available rules.
The rule-based architecture allows clients, who may have defined custom query functions and indexes, to plug in rules for making use of that additional functionality. For example, a client could implement a geospatial index, extend the query API with custom functions for bounding box queries, and write custom rules to plan geospatial filters as scan of the geospatial index, all while making use of built-in rules.

\customparagraph{Future directions.} The experimental planner's IR and rule execution system were designed to anticipate future needs. For example, the data structure used by the ``expression'' intermediate representation currently stores only a single expression at any time: this planner repeatedly rewrites the IR each time a rule is applied. However, it could be seamlessly replaced by the compact Memo data structure~\cite{Graefe95thecascades} which succinctly represents a huge space of possible expressions. The Memo structure replaces each node in the IR tree with a group of logically equivalent expressions. Each group can then be treated as an optimization target where each member of the group represents a possible variant of the group's logical operation. Memo allows optimization work for a small part of the query to be shared (or memoized) across many possible expressions. Adding the Memo structure paves the way to a cost-based optimizer, which uses estimates of a cost metric to choose from several possible plans.

\end{document}